**Title page**

# Contact inhibition of locomotion and junctional mechanics guide collective cell behavior in epithelial wound repair

Luke Coburn[1], Irin-Maya Schouwenaar[4], Hender Lopez[2], Alpha S. Yap[4], Vladimir Lobaskin[3], Guillermo A. Gomez[4]

Affiliations: [1]Institute of Complex Systems and Mathematical Biology, University of Aberdeen, United Kingdom; [2]Centre for BioNano Interactions, School of Chemistry and Chemical Biology, University College Dublin, Belfield, Ireland; [3]School of Physics, University College Dublin, Dublin, Ireland; and [4]Institute for Molecular Bioscience, Division of Cell Biology and Molecular Medicine, The University of Queensland, St. Lucia, Brisbane, Queensland, Australia 4072

*To whom correspondence should be addressed:

Dr. Luke Coburn: luke.coburn@abdn.ac.uk
Dr. Guillermo A. Gomez: g.gomez@uq.edu.au

**Abstract**

Epithelial tissues form physically integrated barriers against the external environment protecting organs from infection and invasion. Within each tissue, epithelial cells respond to different challenges that can potentially compromise tissue integrity. In particular, cells collectively respond by reorganizing their cell-cell junctions and migrating directionally towards the sites of injury. Notwithstanding, the mechanisms that define the spatiotemporal scales and driving forces of these collective responses remain poorly understood. To address this we first analyzed the collective response of epithelial monolayers to injury and compare the results with different computational models of epithelial cells. We found that a model that integrates the mechanics of cells at the cell-cell and cell-substrate interface as well as contact inhibition of locomotion predicts two key properties of epithelial response to injury as: 1) local relaxation of the tissue and 2) collective responses involving the elongation of cells (basal and apical regions) and extension of cryptic lamellipodia that extend up to < 3 cell diameters from the site of injury. Our results therefore highlight the integration between junctional biomechanics, cell substrate adhesion and contact inhibition of locomotion to guide the rapid collective rearrangements that are required to preserve the epithelial barrier in response to injury.

**TOC**

We modeled epithelial cells on their capacity to adhere to one another, to the substrate as well as exhibit contact inhibition of locomotion. We found our model is able to reproduce precisely the main morphological cellular changes and correct spatial scales associated to collective responses required for wound repair.

**Introduction**

Collective cell responses allow epithelial tissues to alter their shape, preserve barrier function and self-repair (Li *et al.*, 2013; Friedl and Mayor, 2017). This is underpinned by the capacity of their constituent epithelial cells to respond to the biochemical and mechanical properties of their surrounding environment (Mammoto *et al.*, 2013; Enyedi and Niethammer, 2015; Lecuit and Yap, 2015; Mao and Baum, 2015). At the sites of cell-cell junctions, adhesion receptors, like cadherins, couple the contractile actomyosin apparatuses of epithelial cells together to generate junctional tension (Yonemura *et al.*, 2010; Borghi *et al.*, 2012; Ratheesh *et al.*, 2012; Conway *et al.*, 2013; Leerberg *et al.*, 2014; Bambardekar *et al.*, 2015). On the other hand, at the cell-substrate interface, integrin receptors interact with ligands in the extracellular matrix and exert forces on these adhesion sites (Grashoff *et al.*, 2010).

An essential property of an epithelial tissue is its capacity to self-repair, which involves important contributions from the neighboring cells at the sites of injury. This has been clearly shown in studies that studied the time course of wound repair following laser-induced epithelial injury (Carayon *et al.*, 1985; Abreu-Blanco *et al.*, 2012; Antunes *et al.*, 2013; Fernandez-Gonzalez and Zallen, 2013). These revealed that neighboring cells up to 5 cell diameters from the site of injury collectively contribute to epithelial healing (Antunes *et al.*, 2013; Lubkov and Bar-Sagi, 2014). However, cells play different roles within this neighborhood, depending on their location. In particular, the cells that immediately border the injured site can form an intra-cellular 'purse string' of acto-myosin that encircles the wound and which, upon contraction, leads to closure of the wound site (or extrusion of dying cells) (Carayon *et al.*, 1985; Abreu-Blanco *et al.*, 2012; Antunes *et al.*, 2013; Fernandez-Gonzalez and Zallen, 2013). However, the purse string mechanism is most important for small wounds. As the area to be repaired becomes larger, surrounding cells shift their behavior and extend more lamellipodia to migrate into the wound area, a process named "lamellipodia crawling" (Abreu-Blanco *et al.*, 2012). Similarly, a transition from "purse string" to "lamellipodial crawling" type of behavior has been shown during epithelial cell extrusion at different monolayer densities (Kocgozlu *et al.*, 2016) suggesting that, in general, changes in epithelial mechanics (larger wounds or crowded epithelia) determine the type of behavior that cells use to preserve barrier integrity.

Irrespective of the precise repair mechanism used, epithelial cells surrounding the wound must collectively remodel their cell-cell junctions, change their morphology as well as reorient themselves in the direction of the injury in order to migrate and cover the damaged area and minimize friction (Tambe *et al.*, 2011). Although different mechanisms have been proposed to underlie this junctional remodeling (Antunes *et al.*, 2013; Hunter *et al.*, 2015) we still do not understand what are the biomechanical principles that govern the spatio-temporal scales of these epithelial collective responses.

To address this question, we have now perform a quantitative morphometric analysis of collective responses to injury and compare these changes with different reduced mechanistic model of epithelial monolayers in which cells adhere to one another and/or to the substrate as well as exhibit contact inhibition of locomotion (CIL, Coburn *et al.*, 2016). By comparing the results of the different models with experimental data we now provide quantitative insights on the forces that drive epithelial collective behavior in response to injury.

**Results**

**Epithelial collective cell responses to injury**

Collective cell behavior is an intrinsic property exhibited by epithelial cells in different contexts such as morphogenesis, wound healing and cancer invasion (Friedl and Mayor, 2017). Several experimental techniques have been used to analyze collective cell behavior in response to injury. Of these, laser micro-irradiation is readily implemented and allows changes in cell morphology and movement to be monitored with high spatio-temporal resolution (Abreu-Blanco *et al.*, 2012; Antunes *et al.*, 2013). Using this methodology we first characterized quantitatively the dynamic changes of epithelial cell shape and collective responses, which will permit direct comparison with results from numerical simulations (see below). Thus, we grew to confluence epithelial MCF-7 cells stably co-expressing a membrane targeted mCherry (MT-mCherry) and nuclear-localized-GFP (NLS-GFP) and then introduced an injury by laser micro-irradiation (Supplementary movie 1). We found that immediately upon an injury, there was a small immediate increase in the apical area occupied by the ~10 cells that were damaged (Figure 1a, b). This initial expansion of the injured area is probably due to a local mechanical relaxation of the tissue since MCF-7 cells exhibit significant amount of junctional tension (Figure 1c). After this local relaxation, the dying cells started to extrude and at the same time neighboring cells started to extend lamellipodia and occupy the substrate previously occupied by the dying cells (Figure 1a, Supplementary movie 1). In addition, we found that as the healing process progressed, epithelial cells immediately adjacent to, and few cell diameters beyond the site of damage, become more elongated with their major axis orientated in the direction of the injury (Figure 1a, 2h).

To quantitatively assess these changes, we then measured different shape descriptors and cell orientations in the direction of the injury as a function of their distance (in cell rows) to the injury site (Figure 2a). First we measured the aspect ratio of cells in the apical region where the zonula adherens of epithelial MCF-7 cells is localized (Gomez *et al.*, 2015) and in the basal region corresponding to the lowest part of the cell-cell interface. Note that due to technical limitations we did not measure cell area at the actual cell-substrate interface, as it is difficult to segment cell boundaries at this location based on the fluorescence of the MT-mCherry marker. Our results show that at time points close to healing, the apical aspect ratio of cells situated between 1 and 3 cell diameters from the injury site increased with respect of their aspect ratio at the time before injury and a similar change was observed in the basal area (Figure 2b).

We then analyzed the relative orientation of cells with respect to the site of injury. We measured this as the angle between the major axis of an ellipse that fits a cell (either in the apical or basal area) and the line that defines the cell's position with respect to the injury site (Figure 2c). We found, again, that for the apical area of cells, this angle becomes smaller (i.e. cells are more aligned, $\Delta \gamma = \gamma_{after} - \gamma_{before} < 0$) as the healing process proceeds. This change (absolute value) is bigger for cells in close proximity to the injury site and decreases to values present before injury for cells located at 4 cells diameters beyond the site of injury (Figure 2d). This resembles the phenomenon of plithotaxis during collective epithelial migration in expanding epithelial cell islands (Zaritsky *et al.*, 2015). In addition, a similar trend was observed in the basal region of cells (Figure 2d). Finally, we also quantitatively measured the skewness of cells in the direction of the injury. In our measurements we refer to skewness to the relative displacement between the basal and apical centroids and this is an index of how "tilted" the cells become as they migrate (Coburn *et al.*, 2016). This parameter ($comp\ \Delta \vec{r}$) compares at some extent to the formation of cryptic lamellipodia that have been observed in collectively migrating cells (Farooqui and Fenteany, 2005; Trepat *et al.*, 2009). In our description (see methods) a "positive" skewness parameter means that cells extend their basal area more than the apical area in the direction of the injury whereas a "negative" means cells extend their basal area in the direction opposite to the injury site (Figure 2e). Our measurements of skewness show that after injury cells preferably extend their basal area in the direction of the injury and this is observed for cells up to 3 cells diameters from the injury site (Figure 2f). Overall, this quantitative data shows that the presence of a micro-injury causes a local relaxation in the tissue and triggers collective responses that involve the elongation of cells and orientation of their major axis in the direction of injury as well as extension of cryptic lamellipodia that extends up to < 3 cell diameters from the site of injury.

**Model of epithelial cells**

The above quantitative data allow us now to investigate the driving forces of collective tissue responses to injury using a mechanistic model epithelial cells based on their capacity to interact with one another and with the substrate (Coburn *et al.*, 2016). Cell-cell adhesion is modeled using a Cellular Potts model (CPM) algorithm (Kabla, 2012; Noppe *et al.*, 2015; Magno *et al.*, 2015; Albert and Schwarz, 2016) where a cell is made up of a given number of pixels that are allowed to change the index (cell ID) that has been assigned to them. The pixel attribution switching is performed according to a probabilistic Metropolis rule, which leads to a minimization of the tissue energy function. The energy function includes terms related to cell-cell adhesion, apical ring contractility, and cell volume as well as cell motility (Kabla, 2012). For cell-substrate adhesion we used our previously described protrusion driven cell propulsion model that incorporates also CIL (Coburn *et al.*, 2013).

Both cell-cell adhesion and cell-substrate adhesion were then coupled by an elastic spring that represents the intracellular level of stiffness as is described in detail in (Coburn *et al.*, 2016). This leads to a modeling approach that, although not strictly three dimensional, allows the analysis in simplified terms of the main properties that define collective epithelial cell behavior in response to injury.

Using this approach we first evaluate the temporal variation of the apical area of epithelial cells using the CPM on a square lattice. In our model, the apical layer of a cell $i$ consists of a region of sites in the lattice with spin $i$ where $i(x, y) = 1, 2, \ldots, N$ and $(x, y)$ gives the lattice position and $N$ is the number of cells in the monolayer. In the initialization of the CPM we choose a resolution $p$ and subdivide the domain into $N$ squares each of side $p$ pixels. Here $p$ is the square root of the preferred area $p = \sqrt{a_p}$. The tissue is modeled with laterally periodic boundary conditions and contiguity is forced, thereby preventing cell partitioning. The energy equation is the same that we used previously (Noppe *et al.*, 2015) with an added apical-basal crosstalk term (Coburn *et al.*, 2016):

$$E_i = K(L_i - \frac{J}{2K})^2 + \lambda(a_i - a_p)^2 + cs|\Delta r(t)|^2 - \frac{J^2}{4K^2} \qquad (1)$$

The first term in (1) accounts for the adhesion between cells and the contractility of the junctions as in (Noppe *et al.*, 2015) and finds a minimum when $L_i = \frac{J}{2K}$ where J determines the strength of cell-cell adhesion and $K$ determines the strength of the junction contractility. The ratio of these two terms therefore determines the preferred boundary length of cells for a monolayer packed at a density $1/a_p$. When this ratio is below the minimum permissible boundary length, which corresponds to the packing of regular hexagons, the system is said to be in the hard regime (Farhadifar *et al.*, 2007; Noppe *et al.*, 2015; Magno *et al.*, 2015; Coburn *et al.*, 2016). The second term accounts for the volume preservation of cells and constraint fluctuations of the apical area at constant height (Coburn *et al.*, 2016). The third term is the apical/basal crosstalk term. As is described in (Coburn *et al.*, 2016), the cell cytoskeleton serves to limit the skew of the columnar cell shape or the lateral displacement ($|\Delta r_i|$) of the xy projection of the apical ($r_i^a$) and basal ($r_i^b$) centroids. Thus, to account for cross-talk between both adhesion systems we assume that the cytoskeleton functions like a spring (with spring constant $cs$) that controls the value of $|\Delta r_i| = |r_i^a - r_i^b|$, always opposing its increase. Finally, the fourth term in (Eq. 1) is constant and does not affect energy changes.

The simulation box size is chosen such that the area of the domain is $Na_p$ so that cells will exactly fill the box thus eliminating internal cell pressure. Then we update the system by randomly selecting pixels at the boundary of cells and changing the value of the pixel's spin (cell ID) to that of its neighbor. The probability (*Prob*) of accepting this change at a given temperature $T$ is given by the energy change of the entire system $\Delta E$ caused by this spin change according to the Metropolis procedure (Graner and Glazier, 1992))

$$Prob\left(i(x,y) \rightarrow i\left(x´,y´\right)\right) = \begin{cases} e^{-\Delta E/T}, & \Delta E > 0 \\ 1, & \Delta E \leq 0 \end{cases} \qquad (2)$$

Then we introduced cell-substrate adhesion in the model by introducing cell protrusion as we described before (Coburn *et al.*, 2013; Coburn *et al.*, 2016). Briefly, a protrusion distribution is assigned to a cell using a discrete set of $m$ points distributed uniformly about the cell center and so the protrusion contour $P$ is defined in polar co-ordinates as

$$P_0(\theta) = A_1 \qquad (3)$$

where $A_1$ is the protrusion radius being the orientation of the cell and regrowth of protrusions calculated as described in (Coburn *et al.*, 2013). In the model cellular protrusions impart a net force $\mathbf{F}_i$ on the cell in the direction of their growth, whose magnitude is proportional to their length (Caballero *et al.*, 2014) according to:

$$\mathbf{F}_i(t) = h_0 \int_0^{2\pi} P_i(\theta, t) \mathbf{n}_\theta \, d\theta \qquad (4)$$

In Eq. 4, $h_0$ is the mobility of the cell and $\mathbf{n}_\theta$ is the unit vector in direction θ. Cell position is then updated using:

$$\frac{dr_i^b(t)}{dt} = [\mathbf{F}_i(t) - s\Delta\mathbf{r}(t)] \qquad (5)$$

To introduce CIL, the region of overlapping protrusions between two adjacent cells is retracted in the radial direction (Coburn *et al.*, 2013; Coburn *et al.*, 2016).

**Numerical simulations**

To analyze the mechanism that drive epithelial collective responses, we performed numerical simulations of epithelial injury in the following three scenarios of our model of epithelial cells:

i. Monolayers of cells with only cell-cell adhesion and junctional contractility.
ii. Monolayers of cells that exhibit adhesion to the substrate and interact with each other trough CIL
iii. A combination of the above reduced models in which cell-cell adhesion is mechanically coupled to cell substrate adhesion through intracellular stiffness.

*i. Injury response in cells that exhibit cell-cell adhesion and junctional contractility but no adhesion to the substrate.* Several models have been developed previously to analyze the local rearrangement of epithelial cells in response to an injury. In particular, computational simulations of apical adherens junctions using vertex or Cellular Potts models have been useful to describe the mechanical behavior of cells that exert low level of traction forces on the substrate (Rozbicki *et al.*, 2015) and how these might respond to injury (Kuipers *et al.*, 2014; Noppe *et al.*, 2015).

To investigate more directly the role of adhesion to the substrate as well as CIL on the collective responses during epithelial repair we first investigated *in silico* the response of the tissue to an injury for monolayers in which cell adhesion to the substrate and CIL was switched off in the model. In our simulations, monolayers were first allowed to equilibrate (i.e. achieve the minimum of the total energy) and then a local injury was introduced in a group of 10 cells by setting as zero their parameters for cell-cell adhesion, junctional contractility and volume conservation, all active processes that cease when a cell dies. Thus, these injured cells are effectively removed from the simulation, which is similar to the removal of dead cells from the epithelial layer that is observed experimentally (Rosenblatt *et al.*, 2001; Kuipers *et al.*, 2014; Lubkov and Bar-Sagi, 2014). As we described before, in the absence of injury this type of modeling lead to steady-state epithelial monolayers in either a hard or a soft regime in which cells exhibit high or low junctional tension, respectively (Noppe *et al.*, 2015; Magno *et al.*, 2015; Coburn *et al.*, 2016). Therefore we investigated the response to injury in these two regimes.

Figure 3 shows results from simulations of the injury response of epithelial cells in the soft and hard regimes. For this series, we keep the adhesion strength fixed ($J = 5875$) and vary only the contractility. We found that for high values of the junctional contractility term ($K \geq 50$, hard regime) the injury area starts growing right after the injury. This is the product of a local mechanical relaxation caused by the presence of injured cells that do not longer contribute to tension generation (Figure. 3a,b). After approx. 30,000 simulation time steps all these curves reach a plateau and show only moderate growth in injury size. The plateau area scales with the junctional contractility parameter ($K$) consistent with the idea that tissue recoil after injury is related to a mechanical relaxation. In contrast, for values of contractility ($K < 50$, soft regime), we observed an immediate decrease of the area occupied by the dying cells after injury (Figure 3a,b). These results show that in the absence of cell-substrate adhesion (and CIL) the relative contribution of the contractility ($K$) and adhesion parameters ($J$) determines whether the injury area will "open" (until it reaches the plateau area) or "close", which agrees with previous computational analysis of wound healing (Kuipers *et al.*, 2014; Noppe *et al.*, 2015).

We then analyzed the presence of collective response in the model by measuring aspect ratios and cell orientation in either soft ($K$=30) or hard regimes ($K \geq 40$). We found in these simulations that during the phase of expansion (hard regime) or contraction (soft regime) of the injured area, cells exhibit a change in their aspect ratio that is higher for cells close to the injury site and decreases to a plateau after only 2 or 3 cell diameters (Fig 3c). Moreover, we found that only cells in direct contact with the injured area (row 1) reorient their principal axis in the direction of the injury (Fig 3d), which contrasts with the experimental observation of a collective re-orientation of cells of up to 3 cell diameters.

Although this model that only incorporates cell-cell adhesion and junctional contractility predicts healing of the epithelial monolayer over long time scales, it is also clear that this model presents two major limitations when compared to experimental results: 1) It has a limited capacity to predict collective morphological changes in regions that surround the injury site, such as the orientation of cells in the direction of the injury. 2) It predicts healing of the tissue in a range of parameters that do not agree with experimental observations, i.e healing occurs in the soft regime only, when cells does not exhibit junctional tension.

*ii. Injury response in confluent monolayers of cells that only interact with the substrate and to each other through CIL.* The results from the previous section motivated us to investigate how cells' traction on their substrate and CIL, (Stramer and Mayor, 2016) contribute to epithelial collective responses in the absence of cell-cell adhesion. Therefore we performed an *in silico* injury in monolayers of cells that lack cell-cell adhesions and junctional tension (Figure 4) and analyzed the responses of the neighboring cells to injury. We found that loss of adhesion to the substrate of the dying cells facilitates cell next to the injury site to extend their protrusions into the injured area (Figure 4a). This allowed a gain of net axial orientation of neighboring cells in the radial direction to the injury site, which permitted these cells to gain traction, migrate and heal the tissue *in silico* (Figure 4b). As expected, we found that healing closes faster for cells with a higher motility parameter ($h_0$, Figure 4b).

One interesting property that we found for this type of *in silico* tissue, compared to the one that only has cell-cell adhesion and not adhesion to the substrate (Figure 3), is that the CIL interaction between the cells next to the injury site guides them to migrate into the free space. As they migrate, free space is opened behind these cells, which promotes the axial orientation of cells in the next row, thus generating a simple mode of collective cell migration that permeates many rows (~3-4 cell diameters) away from the site of injury (Figure 4c). This observation is in line with measurements of the aspect ratio of cells and relative orientation into the direction of the injury (Figure 4d,e, Brugues et al., 2014). Thus, and very simplistically, CIL allows cells to collectively respond to the presence of injury.

*iii. Injury response in monolayers with adhesive and contractile cell-cell junctions, adhesion to the substrate and CIL.* To test how CIL and adhesion to the substrate may contribute to the collective responses to injury of epithelial monolayers in the hard regime we then combined

the above two models: 1) the Cellular-Potts model, to describe cell-cell adhesion and junctional tension and 2) the CIL model, to describe cell substrate anchoring, cell migration and the interaction of cellular protrusions via CIL. In particular, we were interested in analyzing the reaction of cells that are in the hard regime (Figure 5a), in which they exhibit significant amounts of junctional tension (Fig 1b), and whose response to injury was not reproduced by the previous reduced models (Figure 3 and 4). Figure 5b gives the area occupied by dying cells normalized to its pre-injury value vs time for a range of values of the junctional contractility parameter $K$. Strikingly, we found that even for injuries made on monolayers in the hard regime, the apical area occupied by dying cells shows an initial expansion followed by a slow shrinkage, which matches precisely what we observed in our experiments (Figure 1). As before, this initial expansion scales with the junctional contractility (K parameter, Figure 5b), showing that it occurs as a result of the loss of tissue tension due to the removal of the dying cells. Moreover, after this local relaxation of the tissue, the basal protrusions that form in the cells bordering the injured area stop tissue recoil and initiate the healing process by starting to pull surrounding cells into the injury area. Thus, cell-substrate adhesion and CIL in this model favors epithelial repair even for monolayers of cells with high levels of junctional tension.

We now turned our attention to characterize the presence of collective responses within this model by measuring the different shape descriptors and the relative orientation of cells as a function of their position from the site of injury. We found that before healing cells in contact with the injury site extend cellular protrusions and orientate their basal area in the direction of the injury as we determined by measuring the aspect ratio and angle with respect to the direction of injury (Figure 5c, d, Basal). We note however, the magnitude of the changes in aspect ratio in the basal layer were limited as cells increased their junctional contractility (higher $K$ values) and apply a resistive forces to migration (Figure 5c, Basal).

At the apical area we observed a similar trend but that was more sensitive to values of the contractility parameter $K$. In particular, cells in the soft regime ($K$=30, blue line) show collective responses that persist up to 3 cell diameters as evidenced by an decrease (increase) in the magnitude of the cell aspect ratio (relative angle of cells with respect to injury) with the distance from the site of injury (Figure 5,c,d, Apical). A similar trend was observed when we analyzed cell's skewness as a function of their distance from the injury (Figure 5, e), in which this parameter is maximal for cells in contact with the injury and decreases to average at cells beyond to 3 rows from the site of injury. As cells become

"harder, these collective responses still persist but vanishes at K>>60. However, such high contractility might be unrealistic as such value predicts more than 10% elastic recoil, which is more than the elastic relaxation observed experimentally.

Thus, our simulations predict that cells up to 3 cell diameters from the site of injury collectively reorganize, by altering their shape and migration properties to heal the injury. This prediction of the model is in agreement with our experimental results and with previous reports in the literature (Antunes *et al.*, 2013) as well as experimental estimates of length scales of force propagation within tissues (Ng *et al.*, 2014).

**Discussion**

We found that our simple model, combining CIL and junctional mechanics, is sufficient to describe collective morphological cell responses to injury in epithelial tissues. In particular, simulations of monolayers in the hard regime have the capacity of exhibit salient properties of epithelial tissues that have been observed experimentally: i) cells exhibit junctional tension, ii) after injury there is a local relaxation of the tissue, iii) the tissue is able to heal and iv) healing involves morphological collective responses of epithelial cells in the neighborhood of the site of injury.

Results from our simulations and experiments show how cell-substrate adhesion and CIL play crucial roles in the injury response. First, we observed that an initial expansion of the area of injury observed in simulations (and in experiments) is a consequence of epithelial cells' ability to generate junctional tension. Loss of junctional tension caused by the injury leads to an elastic mechanical relaxation of the cells surrounding the injured area. However, the extent to which this local relaxation is propagated across the tissue is limited by the cell adhesion to the substrate, similarly to what is observed in epithelial cell islands (Ng *et al.*, 2014; Coburn *et al.*, 2016). Second, cell adhesion to the substrate and CIL allow the extension of protrusions and the migration of cells into the injury area. During this process cells at the edge of the injured area migrate first leaving gaps behind that favors the orientation, asymmetry and migration of the cells thus generating a collective response to injury. Of note, these collective rearrangements are not observed in simulations in which cells are only allowed to interact with each other and generate junctional contractility (Figure 3). In addition, our results also show that extension of protrusions both *in vivo* (Figure 1, Supplementary Movie 1) and *in silico* (Figure 5) occurs later and on a timescale that is slower than the instantaneous elastic relaxation, being this similar for a wide range of parameter values (Figure 5). This reveals the robustness of this model to reproduce these spatio-temporal features of the epithelial response to injury.

Another important aspect of our model is that also provides an explanation for the length scales at which collective responses originate in response to injury (Farooqui and Fenteany, 2005; Antunes *et al.*, 2013; Lubkov and Bar-Sagi, 2014). In particular mechanical coupling between cell-cell adhesion and cell-substrate adhesion with CIL serves as a mechanism that leads to a stress gradient that propagates to cells behind the sites of injury altering their shape and polarizing them collectively in the direction of the injury. A similar stress gradient has been reported for collective behavior in expanding epithelial islands (Banerjee *et al.*, 2015; Zimmermann *et al.*, 2016), thus suggesting that the biomechanical properties of cells profoundly affects the spatial scales of collective responses.

Finally, a number of candidate active responses that drive collective behavior of cells in response to injury have been described in the literature. For example, neighboring cells may soften their bonds to facilitate cell rearrangements that occur during neural tube closure and cell extrusion (Escuin *et al.*, 2015; Hashimoto *et al.*, 2015); observations that agreed with our numerical simulations of "soft" cells (Figure 5). In addition, directional flow of actomyosin towards the injury-live cell interface to form an actomyosin purse string could be favored by the orientation of protrusions and changes in cell aspect ratios (Antunes *et al.*, 2013), a notion supported by the work of the Ladoux's Lab where it is shown that purse strings are preferentially stabilized in the regions of negative curvature (Ravasio *et al.*, 2015). Finally, we do not exclude the possibility that motility may be enhanced, through mechanotransduction in the tissue next to injury site. Cells next to the injury site gain greater traction to the substrate due to a passive resistance to migration from the cells in the next rows (Coburn *et al.*, 2016). This could in principle enhance cell's traction on the substrate through force transduction as it has been observed in other experimental systems (Weber *et al.*, 2012; Mertz *et al.*, 2013). Although all of these phenomena have been shown to contribute, to some extent, to the collective response of neighboring cells to the sites of injury, there is no clear consensus on what limits the spatio-temporal scales of this process. Our results show that the presence of CIL and the ability of cells to adhere to the substrate allow cells to extend their basal area relatively more than the apical area thus polarizing them towards the site of injury as well as limiting the space over which mechanical relaxation of the tissue occurs. We hypothesize that these mechanisms allow cells to transmit information about the presence and location of the injury across the tissue that can be used by other cells to generated controlled and local active responses as proliferation (Aragona *et al.*, 2013).

**Materials and methods**

*Cell Culture and Transfections*

MCF-7 cells were from ATCC and cultured in DMEM; supplemented with 10% foetal bovine serum (FBS), 1% non-essential amino acids, 1% L-glutamine, 100 U/ml penicillin and 100 U/ml streptomycin. Cells were infected with lentivirus expressing mCherry–K-Ras$^{C14}$ (mCherry-MT) and Hist2b-GFP (NLS-GFP) as it was described in (Leerberg *et al.*, 2014; Wu *et al.*, 2014) and mCherry and GFP positive cells were isolated by dual colour fluorescent activated cell sorted and subsequently maintained in DMEM+10% FBS plus antibiotics as described above.

*Two-photon laser induced epithelial injury*

Laser microirradiation was performed as described in (Michael *et al.*, 2016). Confluent cells stably expressing a plasma membrane targeted mCherry construct (mCherry–K-Ras$^{C14}$, Wu *et al.*, 2014). A 354 x 354 µm region was imaged at 90 sec intervals for ~5 hours with a total of 11 z-slices (1µm thick). To induce cell injury, a circular 85 µm diameter (~10 cells) centered in the field of view was irradiated at the vertical center of the monolayer (z-slice position 6) after one frame of starting the acquisition. Injury was carried out using 35% transmission of the 790nm laser for 35 iterations.

*Laser ablation experiments to measure junctional tension.*

The use of laser ablation technique to assess junctional tension has been described in detail previously (Liang *et al.*, 2016). Briefly, cells stably expressing E-cadherin-GFP in an E-cadherin shRNA knockdown background (Smutny *et al.*, 2011; Priya and Gomez, 2013) were used to identify the apical region of cell-cell contacts. These experiments were carried out at 37$^0$C on a Zeiss LSM710 system (63x, 1.4NA Plan Apo objective) using a 488 nm laser for time lapse imaging (GFP and DIC imaging) and a MaiTai (Coherent) laser set at 28% transmission and 790nm for ablation. Time lapse imaging for ~3 min (20 frames) of a 90 x 90 µm region was taken at 2 sec intervals and ablation was done after the second frame of acquisition on a 2 µm diameter circular region on the apical cell-cell junctions. Values shown average recoil curves for 15 ablated contacts.

*Image analysis.*

*Injury area over time.* The area of injury and its variation were determined in Image J by drawing a region of interest (ROI) corresponding to the damaged area in the plane of the monolayer and following its changes over time. Data were normalized to the area occupied by cells that died upon injury before ablation. Data shown correspond to the average of 3 independent movies.

*Collective cell rearrangements.* First, cells were indexed with respect to their position (in row number) to the site of injury. Then, ROIs corresponding to the boundaries of each cell in its most apical and most basal planes were drawn using the drawing tools in Image J and added to the ROI manager. In addition, an additional ROI that describes the region of injury (or region occupied by dying cells) was drawn. After ROIs were drawn the following options were chosen in the set measurements menu in Image J i) centroid, ii) shape descriptors and iii) fit ellipse. Thus, after "measure" in Image J, we extracted the following parameters from each ROI: a) X and Y coordinates of the centroid of each ROI, b) the angle between the major axis of an ellipse that fits the cell boundary and the X axis of the image and 3) the aspect ratio (ratio between the major axis and minor axis of the ellipse that fits the cell boundary). Using this information we then calculated the average value of the following quantities for each cell row from the site of injury:

i. *Change of Aspect ratio before and after injury:* We measured for each cell how much their aspect ratio changes before and after injury ($t$=2 h). These measurements were done for both the basal and apical region of cells. Average values were then calculated for cells in the same row within a movie.
ii. *Change of cells' orientation with respect to the site injury before and after injury*. For these measurements we first calculated the vector position of a cell with respect to the site of injury using the information of the cells centroid and the centroid of the area of injury. From this information, we calculated the orientation (or slope, m1) of the line that connects the site of the injury and the analyzed cell. Similarly, by fitting a cell boundary with an ellipse we calculated the orientation (or slope, m2) of the major axis of a cell with respect of an image using the "angle" values obtained from image J. Using the values of m1 and m2 we then calculated the acute angle ($\gamma$) between cells orientation and the direction to the injury using

$$\tan \gamma = \left|\frac{m_1 - m_2}{1 + m_1 m_2}\right| \quad (6)$$

and how it changes before and after (2 h) injury (i.e. $\Delta\gamma = \gamma_{after} - \gamma_{before}$).

iii. Changes in cell skewness before and after injury. A measure on how much a cell is tilted within the monolayer is given by the 2D projection in the basal plane of the vector that connect the basal and apical centroids (Coburn *et al.*, 2016) and a measure on how much this is aligned in the direction of the injury is given by the component of this vector in the direction that connect the basal centroid of the cell and the centroid of the area of injury. Thus we measured the magnitude of this component ($comp.\overrightarrow{\Delta r}$) using the dot product equation between these two vectors and dividing it by the distance between the injury site (centroid) and the position of the basal centroid of the analyzed cell. Average changes in $comp.\overrightarrow{\Delta r}$ before and after 2 h of injury were calculated for each cell row.

Results shown in figures correspond to average values obtained from 3 independent movies.

*Code availability*

The codes used for numerical simulations are available upon request.

**Acknowledgments**

We thank all our lab colleagues for their support and advice. This work was supported by grants from the National Health and Medical Research Council of Australia (1067405, 1123816 to AY and GAG; 1037320 to AY). ASY is a Research Fellow of the NHMRC (1044041). LC was funded under the Higher Education Authority of Ireland's Programme for Research in Third Level Institutions (PRTLI) Cycle 5 "Simulation Science: and co-funded by the European Regional Development Fund (ERDF). H.L. would like to acknowledge the financial support of the Irish Research Council, Enterprise Partnership Scheme Postdoctoral Fellowship Programme (Project ID EPSPD/2015/5). GAG is supported by an Australian Research Council Future Fellowship (FT160100366). Optical imaging was performed at the ACRF/IMB Cancer Biology Imaging Facility, established with the generous support of the Australian Cancer Research Foundation.

## Figure legends

**Figure 1. Epithelial collective rearrangements in response to injury**. **A)** Confluent MCF-7 cells expressing a plasma membrane targeted mCherry (MT-Cherry) were locally injured by micro-irradiation (dashed yellow line) with a two-photon laser and the morphological responses were analyzed by 3D (z-stack) time lapse imaging. Panels show merged projections of the plane of cells in contact to the substrate (basal, pseudocoloured in magenta) and the plane of cells that contain the cell-cell junctions (apical, pseudo-coloured green) at times before injury and at 0 min, 25 min and 2 h after injury. Scale bar 50 μm. **B)** Quantifications of the area within the monolayer occupied by injured cells as a function of time. Plot shows the mean± S.E.M of 3 independent experiments. **C)** Average recoil of the Zonula Adherens after ablation. Experiments were performed on cells expressing E-cadherin-GFP as described in materials and methods. Data show the average recoil for 16 analyzed junctions and its SEM.

**Figure 2. Collective morphological responses of epithelial cells in response to injury.** **A)** Definition of cell positions with respect to the injury center used for quantitative analysis. **B)** Changes in aspect ratio at the apical and basal plane of cells (Δ=Final-initial, see also Materials and methods) in response to injury. Changes were measured as a function of the distance of cells from the site of injury. **C and D)** Scheme (C) and quantification (D) of the changes in relative orientation of cells (measured either at its apical or basal plane, see Materials and methods) as a function of their distance from the site of injury. **E and F).** Scheme (E) and quantitation (F) of the component (comp.) of the vector ($\Delta \vec{r}$) that defines the offset between apical ($r_a$) and basal ($r_b$) centroids in the direction of the injury.

**Figure 3. Epithelial cell response to injury in the absence of cell-substrate adhesion.** **A)** Snapshot of simulations for two values of the contractility parameter K=30 (soft), K=80 (hard) at t=0, t=1.2 x10$^4$ and t=9x10$^4$ Monte Carlo cycles. B) Dynamics of the relative injury area for monolayers for different values of the contractility parameter K. For all these simulations the cell-cell adhesion *J* parameter was kept constant and equal to 5875. **C and D)** Aspect ratio (C) and relative orientation with respect to injury (γ angle, D) of cells in response to injury. Changes were measured as a function of the distance (in cell rows) from the site of injury.

**Figure 4. CIL lead to collective response of the cell-substrate interface in the absence of cell-cell adhesion and junctional contractility. A)** Snapshot of simulations at different time points after injury. **B)** Dynamics of the relative injury area for monolayers with different values of the motility parameter $h_0$. **C)** High magnification (from A) showing the collective alignment of cells in the direction of the injury. **D-E)** Aspect ratio (D) and relative orientation with respect to injury (γ angle, E) of cells in response to injury.

**Figure 5. CIL and junctional mechanics guides epithelial collective cell behavior in a mechanically tense epithelia.**

**A)** Snapshot of simulations for two values of the contractility parameter K=30 (soft) and K=60 (hard) at t=0, t=1.2 x$10^4$ and t=9x$10^4$ Monte Carlo cycles. **B)** Dynamics of the relative injury area for monolayers for different values of the contractility parameter *K*. For all these simulations the cell-cell adhesion (*J)* and motility ($h_0$) parameters were kept constant and equal to 5875 and 0.05, respectively. **C and D)** Aspect ratio (C) and relative orientation with respect to injury (γ angle, D) of cells in response to injury measured both at their apical and basal regions. **E)** Quantitation of the component of the vector that defines the offset between apical ($r_a$) and basal ($r_b$) centroids (comp. $\Delta\vec{r}$ ) in the direction of the injury .

**Supplementary Movie 1**. Confluent MCF-7 cells expressing a plasma membrane targeted mCherry (MT-Cherry) were locally injured by laser micro-irradiation with a two-photon laser and the morphological responses were analyzed by 3D (z-stack) time-lapse (30 s interval) confocal microscopy. Images show merged projections of the plane of cells in contact to the substrate (basal, magenta pseudo-color) and the plane of cells that contain the cell-cell junctions (apical, green pseudo-color).

# Figure 1

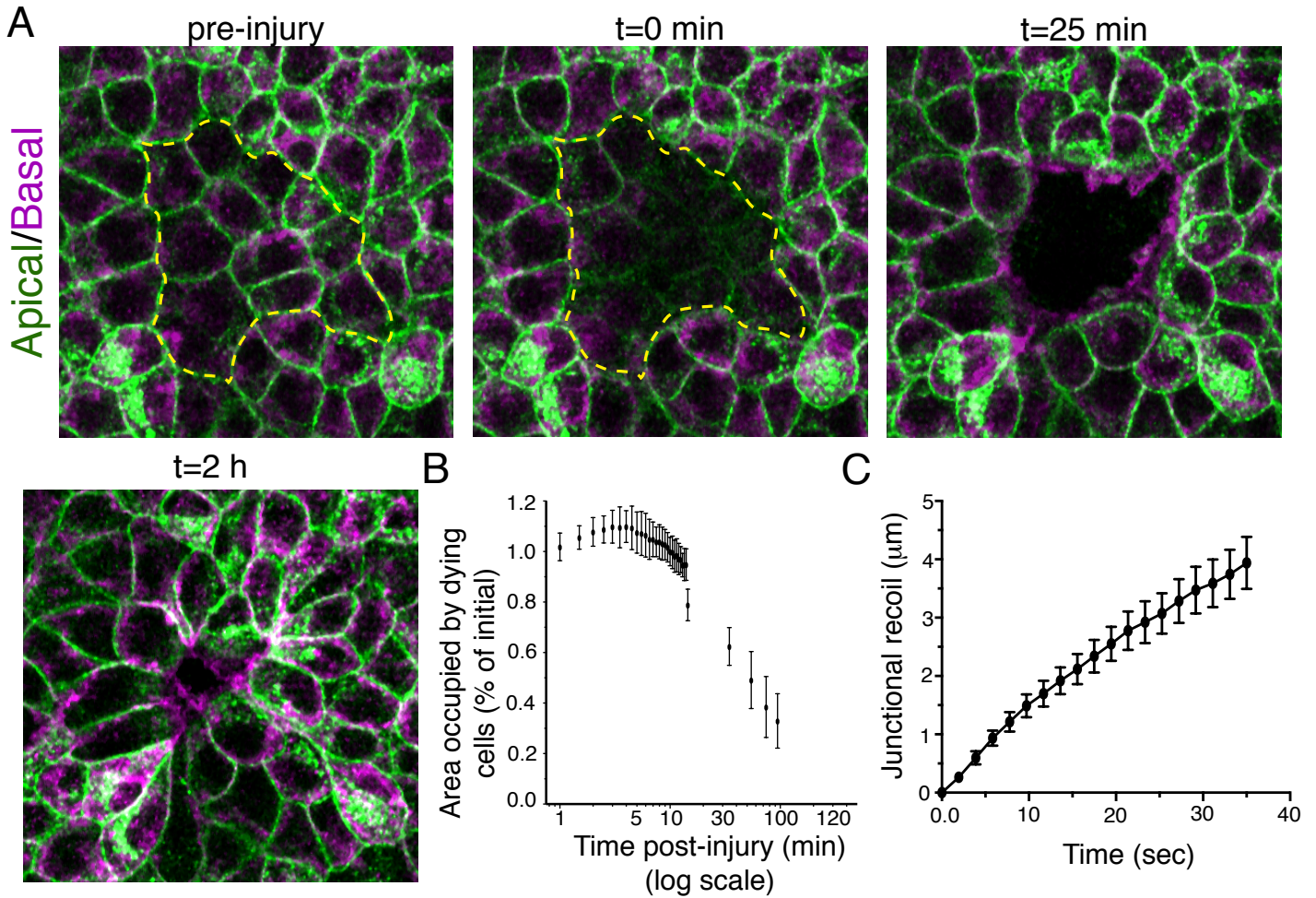

**Figure 1. Epithelial collective rearrangements in response to injury**. **A)** Confluent MCF-7 cells expressing a plasma membrane targeted mCherry (MT-Cherry) were locally injured by micro-irradiation (dashed yellow line) with a two-photon laser and the morphological responses were analyzed by 3D (z-stack) time lapse imaging. Panels show merged projections of the plane of cells in contact to the substrate (basal, pseudocoloured in magenta) and the plane of cells that contain the cell-cell junctions (apical, pseudo-coloured green) at times before injury and at 0 min, 25 min and 2 h after injury. Scale bar 50 μm. **B)** Quantifications of the area within the monolayer occupied by injured cells as a function of time. Plot shows the mean± S.E.M of 3 independent experiments. **C)** Average recoil of the Zonula Adherens after ablation. Experiments were performed on cells expressing E-cadherin-GFP as described in materials and methods. Data show the average recoil for 16 analyzed junctions and its SEM.

# Figure 2

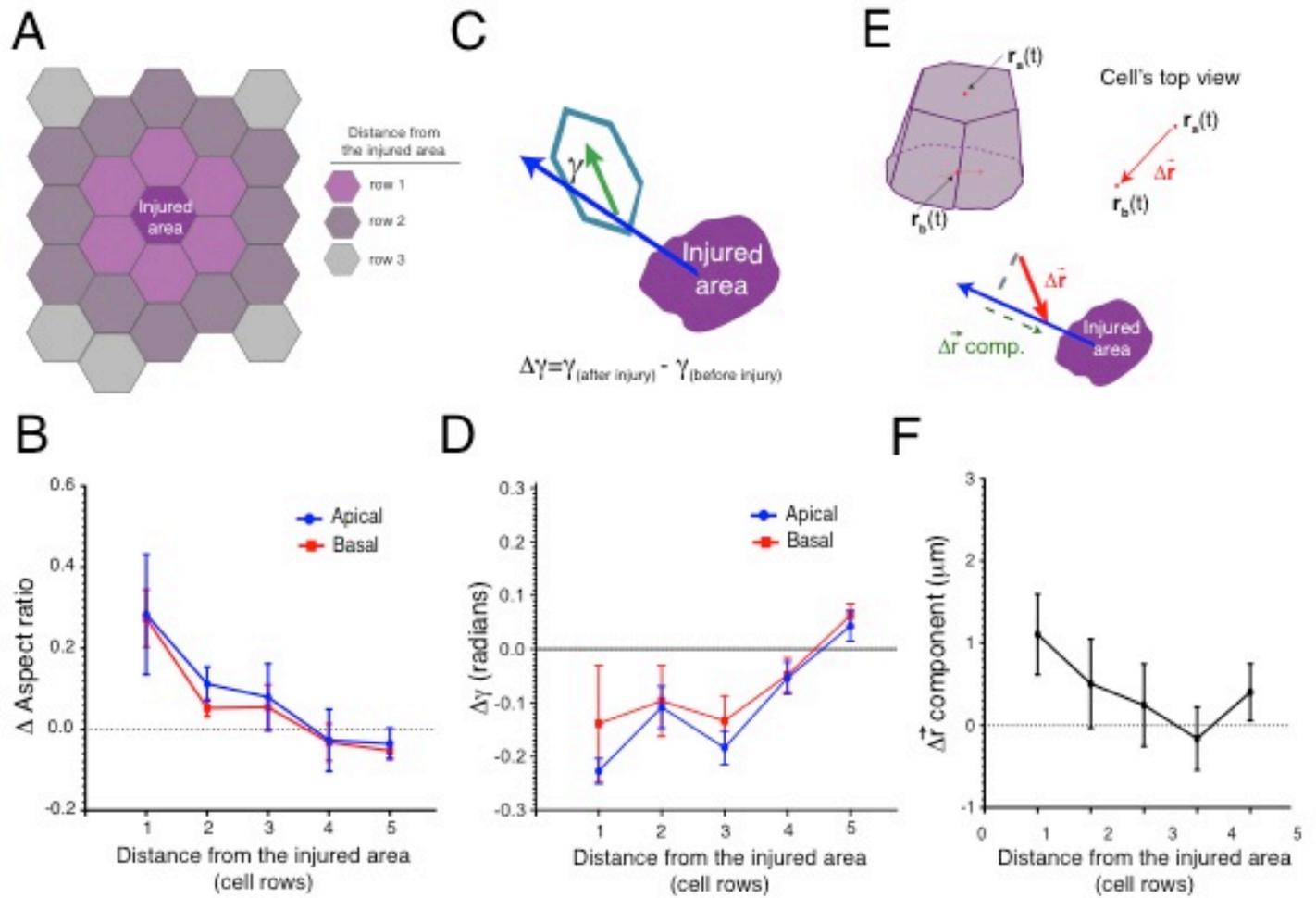

**Figure 2. Collective morphological responses of epithelial cells in response to injury. A)** Definition of cell positions with respect to the injury center used for quantitative analysis. **B)** Changes in aspect ratio at the apical and basal plane of cells (Δ=Final-initial, see also Materials and methods) in response to injury. Changes were measured as a function of the distance of cells from the site of injury. **C and D)** Scheme (C) and quantification (D) of the changes in relative orientation of cells (measured either at its apical or basal plane, see Materials and methods) as a function of their distance from the site of injury. **E and F).** Scheme (E) and quantitation (F) of the component (comp.) of the vector ($\Delta \vec{r}$) that defines the offset between apical ($r_a$) and basal ($r_b$) centroids in the direction of the injury.

# Figure 3

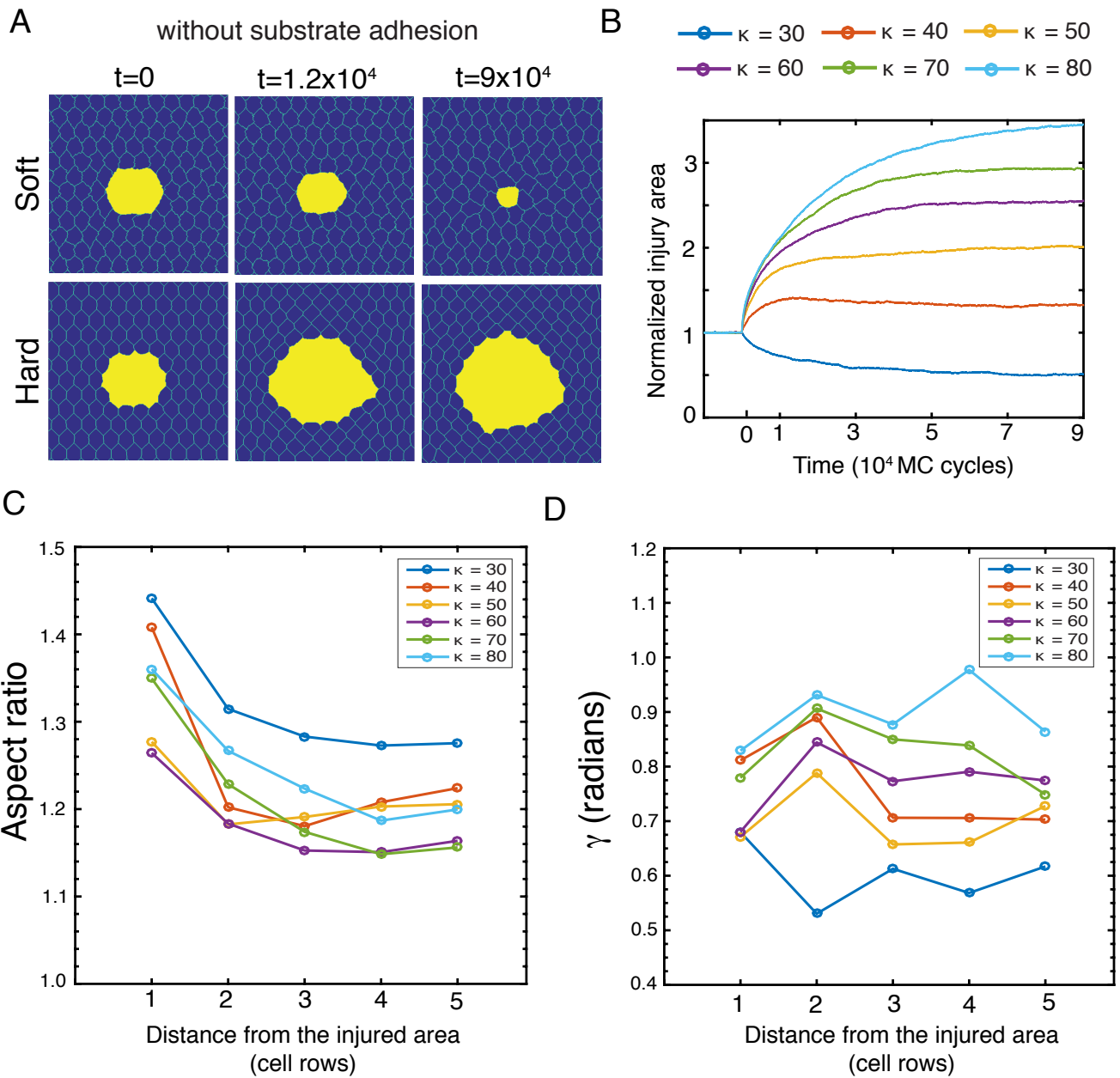

**Figure 3. Epithelial cell response to injury in the absence of cell-substrate adhesion. A)** Snapshot of simulations for two values of the contractility parameter K=30 (soft) and K=80 (hard) at t=0, t=1.2 x10$^4$ and t=9x10$^4$ Monte Carlo cycles. B) Dynamics of the relative injury area for monolayers for different values of the contractility parameter K. For all these simulations the cell-cell adhesion J parameter was kept constant and equal to 5875. **C and D)** Aspect ratio (C) and relative orientation with respect to injury (γ angle, D) of cells in response to injury. Changes were measured as a function of the distance (in cell rows) from the site of injury.

# Figure 4

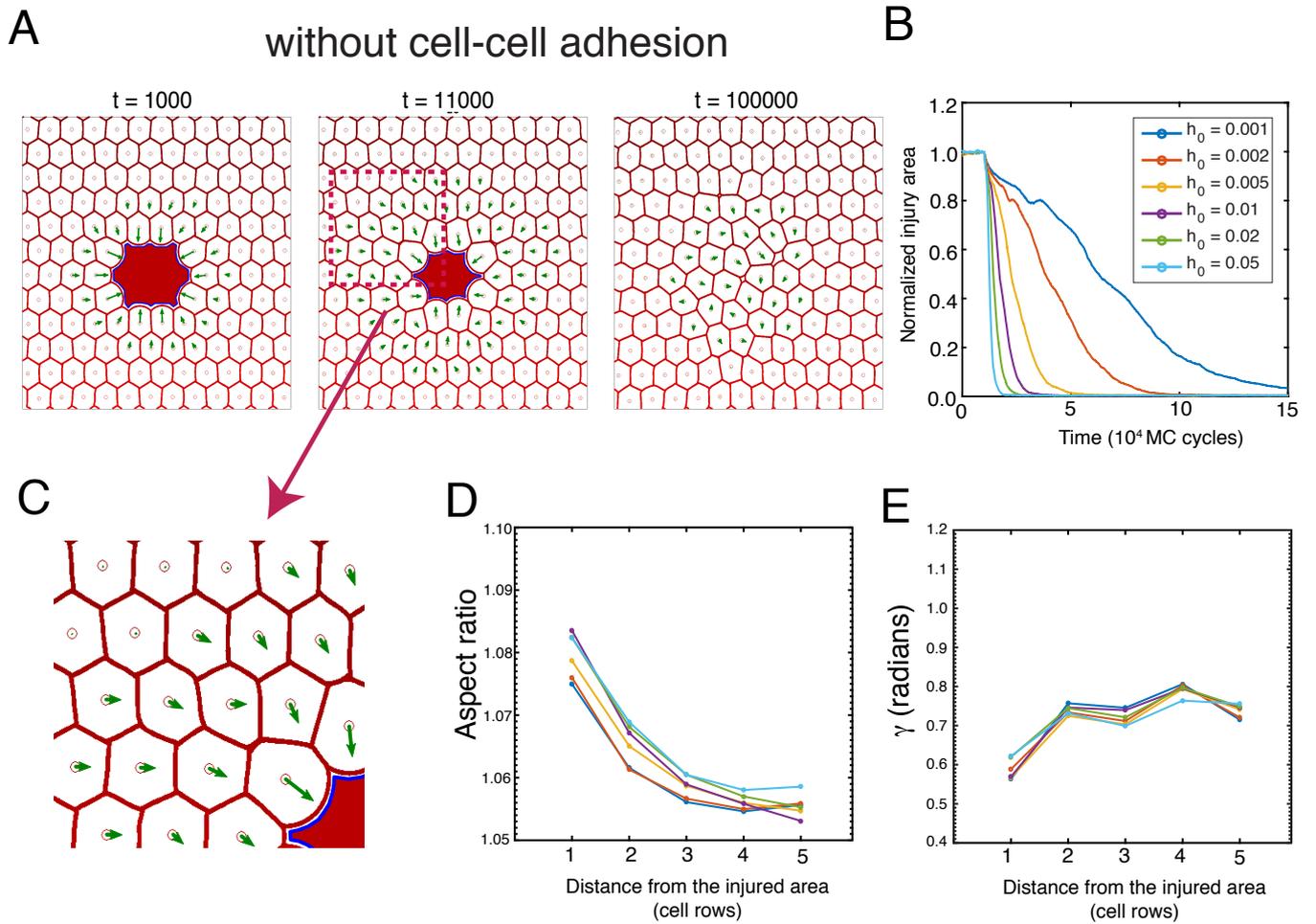

**Figure 4. CIL lead to collective response of the cell-substrate interface in the absence of cell-cell adhesion and junctional contractility. A)** Snapshot of simulations at different time points after injury. **B)** Dynamics of the relative injury area for monolayers with different values of the motility parameter $h_0$. **C)** High magnification (from A) showing the collective alignment of cells in the direction of the injury. **D-E)** Aspect ratio (D) and relative orientation with respect to injury ($\gamma$ angle, E) of cells in response to injury.

# Figure 5

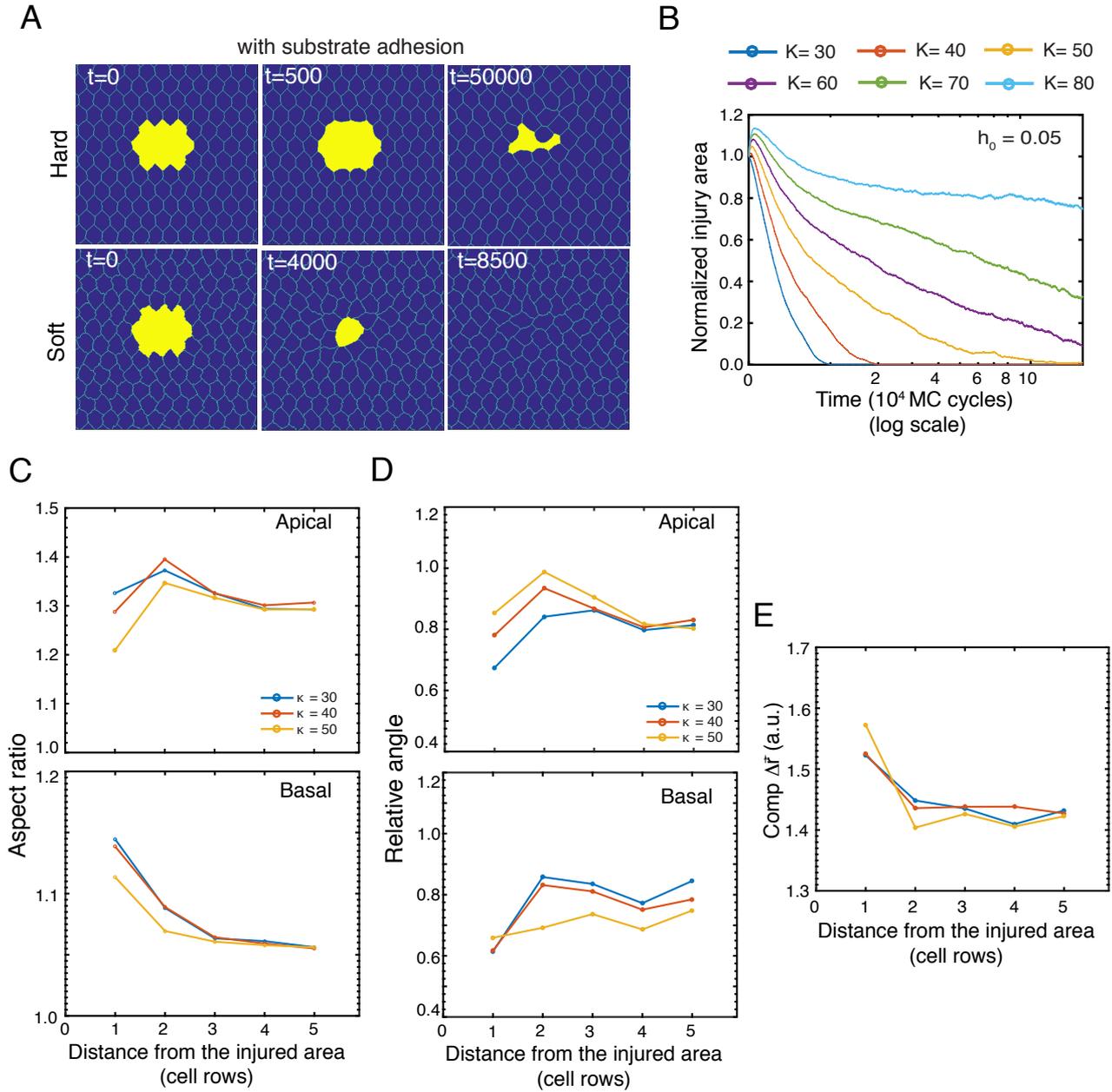

**Figure 5. CIL and junctional mechanics guides epithelial collective cell behavior in a mechanically tense epithelia. A)** Snapshot of simulations for two values of the contractility parameter K=30 (soft) and K=60 (hard) at t=0, t=1.2 x10$^4$ and t=9x10$^4$ Monte Carlo cycles. **B)** Dynamics of the relative injury area for monolayers for different values of the contractility parameter K. For all these simulations the cell-cell adhesion (*J*) and motility (h$_0$) parameters were kept constant and equal to 5875 and 0.05, respectively. **C and D)** Aspect ratio (C) and relative orientation with respect to injury (γ angle, D) of cells in response to injury measured both at their apical and basal regions. **E)** Quantitation of the component of the vector that defines the offset between apical ($r_a$) and basal ($r_b$) centroids (comp. $\Delta \vec{r}$ ) in the direction of the injury.